\documentclass[useAMS,usegraphicx,]{aa}
\usepackage{color}
\newcommand{\xmm}{{\em XMM-Newton~}}
\newcommand{\xmmv}{{\em XMM-Newton,~}}
\newcommand{\nustar}{{\em NuSTAR~}}
\newcommand{\chandra}{{\em Chandra~}}
\newcommand{\integral}{{\em INTEGRAL~}}

\usepackage[space]{grffile}
\usepackage[]{amsmath}
\usepackage[normalem]{ulem}
\usepackage{natbib}
\usepackage{hyperref}
\bibpunct{(}{)}{;}{a}{}{,}
\defcitealias{kaastra2014science}{K14}
\defcitealias{prova}{M14}
\begin{document}

\title{Anatomy of the AGN in NGC 5548:\\III. The high-energy view with NuSTAR and INTEGRAL}
\author{F. Ursini\inst{\ref{ujf}}\fnmsep\inst{\ref{ipag}}\fnmsep\inst{\ref{roma3}}
		\and R. Boissay\inst{\ref{geneva}}
		\and P.-O. Petrucci\inst{\ref{ujf}}\fnmsep\inst{\ref{ipag}}
		\and G. Matt\inst{\ref{roma3}}
		\and M. Cappi\inst{\ref{inafbo}}
		\and S. Bianchi\inst{\ref{roma3}}
		\and J. Kaastra\inst{\ref{sron}}
		\and F. Harrison\inst{\ref{caltech}}
		\and D.~J. Walton\inst{\ref{jpl}}\fnmsep\inst{\ref{caltech}}
		\and L. di Gesu\inst{\ref{sron}}
		\and E. Costantini\inst{\ref{sron}}
		\and B. De Marco\inst{\ref{mplanck}}
		\and G. A. Kriss\inst{\ref{stsi}}\fnmsep\inst{\ref{jhop}}
		\and M. Mehdipour\inst{\ref{sron}}\fnmsep\inst{\ref{mullard}}
		\and S. Paltani\inst{\ref{geneva}}
		\and B. M. Peterson\inst{\ref{ohio1}}\fnmsep\inst{\ref{ohio2}}
		\and G. Ponti\inst{\ref{mplanck}}
		\and K. C. Steenbrugge\inst{\ref{iachile}}\fnmsep\inst{\ref{oxford}}
          }

   \institute{
   			 {Univ. Grenoble Alpes, IPAG, F-38000 Grenoble, France.\label{ujf}} \\
              e-mail: \href{mailto:francesco.ursini@obs.ujf-grenoble.fr}{\texttt{francesco.ursini@obs.ujf-grenoble.fr}}
    \and CNRS, IPAG, F-38000 Grenoble, France. \label{ipag}
    	\and Dipartimento di Matematica e Fisica, Università degli Studi Roma Tre, via della Vasca Navale 84, 00146 Roma, Italy. \label{roma3}
	\and Department of Astronomy, University of Geneva, 16 Ch. d’Ecogia, 1290 Versoix, Switzerland. \label{geneva}
	\and INAF-IASF Bologna, Via Gobetti 101, I-40129 Bologna, Italy. \label{inafbo}
	\and SRON Netherlands Institute for Space Research, Sorbonnelaan 2, 3584 CA Utrecht, the Netherlands. \label{sron}
	\and Cahill Center for Astronomy and Astrophysics, California Institute of Technology, Pasadena, CA 91125, USA. \label{caltech}
	\and Jet Propulsion Laboratory, 4800 Oak Grove Dr., Pasadena, CA 91109, USA. \label{jpl}
		\and {Max-Planck-Institut für extraterrestrische Physik, Giessenbachstrasse, D-85748 Garching, Germany.\label{mplanck}}
	 \and {Space Telescope Science Institute, 3700 San Martin Drive, Baltimore, MD 21218, USA.\label{stsi}}
	\and {Department of Physics and Astronomy, The Johns Hopkins University, Baltimore, MD 21218, USA.\label{jhop}}
		\and {Mullard Space Science Laboratory, University College London, Holmbury St. Mary, Dorking, Surrey, RH5 6NT, UK.\label{mullard}}
			\and {Department of Astronomy, The Ohio State University, 140 W 18th Avenue, Columbus, OH 43210, USA.\label{ohio1}}
	\and {Center for Cosmology \& AstroParticle Physics, The Ohio State University, 191 West Woodruff Avenue, Columbus, OH 43210, USA.\label{ohio2}}
	\and I{nstituto de Astronomía, Universidad Católica del Norte, Avenida Angamos 0610, Casilla 1280, Antofagasta, Chile.\label{iachile}}
	\and {Department of Physics, University of Oxford, Keble Road, Oxford, OX1 3RH, UK.\label{oxford}}
	}

\date{Released Xxxx Xxxxx XX}


\label{firstpage}

\abstract{
We describe the analysis of the seven broad-band X-ray continuum observations of the
archetypal Seyfert 1 galaxy NGC~5548 that were obtained with {\it
XMM-Newton} or {\it Chandra}, simultaneously with high-energy ($> 10$\,keV) observations with {\it NuSTAR} and
{\it INTEGRAL}. These data were obtained as part of a multiwavelength
campaign undertaken from the summer of 2013 till early 2014. We find evidence of a
high-energy cut-off in at least one observation, which we attribute to
thermal Comptonization, and a constant reflected component that is
likely due to neutral material at least a few light months away from
the continuum source. We confirm the presence of strong, partial
covering X-ray absorption as the explanation for the sharp decrease in
flux through the soft X-ray band.  The obscurers appear to be variable
in column density and covering fraction on time scales as short as
weeks. A fit of the average spectrum over the range 0.3~--~400 keV
with a realistic Comptonization model indicates the presence of a hot
corona with a temperature of $40^{+40}_{-10}$ keV and an optical
depth of $2.7^{+0.7}_{-1.2}$ if a spherical geometry is assumed.}
\maketitle
\section{Introduction}
The high-energy spectrum of active galactic nuclei (AGNs) is typically dominated by a primary power law component, which is generally believed to originate in a hot plasma, the so-called corona, localized in the inner part of the accretion flow \citep{reis&miller2013}. The presence of a high-energy cut-off around $\sim$100 keV has been observed in several sources \citep[see, e.g.,][]{perola2002,2041-8205-782-2-L25,IC4329A_Brenneman,marinucci2014swift,ballantyne2014} and is commonly interpreted as the signature of thermal Comptonization. Namely, the optical/UV photons from the accretion disk are upscattered by the hot electrons of the corona up to X-ray energies \citep[see, e.g.,][]{haardt&maraschi1991,hmg1994,hmg1997}. This primary emission can be modified by different processes, such as absorption from surrounding gas and Compton reflection from the disk itself or more distant material. 
Moreover, the spectra of many AGNs rise smoothly below 1-2 keV above the extrapolated high-energy power law \citep[see, e.g.,][]{caixa1}. The nature of this so-called soft excess is still under debate \citep[see, e.g.,][]{done2012SE}.\\ \\
The geometrical and physical properties of the different structures leading to the different spectral components are still poorly known. First and foremost, a more complete understanding of the high-energy emission of AGNs requires disentangling the spectral properties of each component and constraining their characteristic parameters. For this goal, the availability of multiple broad-band observations with a high signal-to-noise ratio is essential. Using the data from a previous long multiwavelength campaign on \object{Mrk 509} \citep{kaastra2011mrk509}, \cite{pop2013mrk509} made a detailed study of the high-energy spectrum of that AGN. In that case, the simultaneous data covering the range from optical/UV to gamma-ray energies made it possible to test Comptonization models in full detail and provide a physically motivated picture of the inner regions of the source. The hard X-ray continuum was well fitted by a Comptonized spectrum produced by a hot, optically thin corona with $kT_e \sim 100$ keV and $\tau \sim 0.5$, probably localized in the very inner part of the accretion flow \citep{pop2013mrk509}. \cite{mehdipour2011509}, studying the optical to X-ray variability, found a correlation between the optical/UV and soft (<0.5 keV) X-ray flux, but no correlation between the optical/UV and hard (>3 keV) X-ray flux. This result was interpreted as indicating the presence of a second, "warm" corona with $kT_e \sim 1$ keV and $\tau \sim 15$, responsible for the soft excess as well as the optical/UV emission \citep{mehdipour2011509,pop2013mrk509}. \\ \\
In the present paper, we discuss results based on a multiwavelength monitoring program on the AGN \object{NGC~5548}, an X-ray bright Seyfert 1 galaxy, hosting a supermassive black hole of $3.9 \times 10^7$ solar masses \citep{pancoast2013mbh5548} and lying at a redshift of $z=0.017175$ \citep[][ as given in the NASA/IPAC Extragalactic Database]{devac1991}. The source historically presented a flux variability typical of an unobscured AGN of that black hole mass \citep{caixa3}. High spectral resolution X-ray and UV observations of this source have previously shown a typical ionized outflow, seen as the "warm absorber" through narrow lines absorbing the UV and soft X-ray emission from the core \citep{crenshaw2003wa5548,steenbrugge2005wa5548}.\\ \\
The multiwavelength campaign was conducted from May 2013 to February 2014, involving observations with the \textit{XMM-Newton}, \textit{HST}, \textit{Swift}, \textit{NuSTAR}, \textit{INTEGRAL} and \textit{Chandra} satellites \citep[][hereafter K14]{kaastra2014science}. During the campaign, the nucleus was found to be obscured by a long-lasting and clumpy stream of ionized gas never seen before in this source. This heavy absorption manifests itself as a 90\% decrease of the soft X-ray continuum emission, associated with deep and broad UV absorption troughs. The obscuring material has a distance of only a few light days from the nucleus, and it is outflowing with a velocity up to five times faster than those observed in the warm absorber \citepalias{kaastra2014science}. Given its properties, this new obscuring stream likely arises in the accretion disk.\\
This work is part of a series of papers focusing on different aspects of the campaign. Here we concentrate on the broad band (0.3-400 keV) X-ray emission, focusing on the observations featuring high-energy ($>$10 keV) data provided by \nustar and \textit{INTEGRAL}.\\ \\ 
The paper is organized as follows. In Sect. \ref{sec:obs&data}, we describe the observations and data reduction. In Sect. \ref{sec:analysis} we present the analysis of the hard (>5 keV) X-ray and total (0.3-400 keV) X-ray spectrum, fitted with both a phenomenological and a realistic Comptonization model. In Sect. \ref{sec:discussion}, we discuss the results and summarize our conclusions.
\begin{figure}
\includegraphics[width=\linewidth]{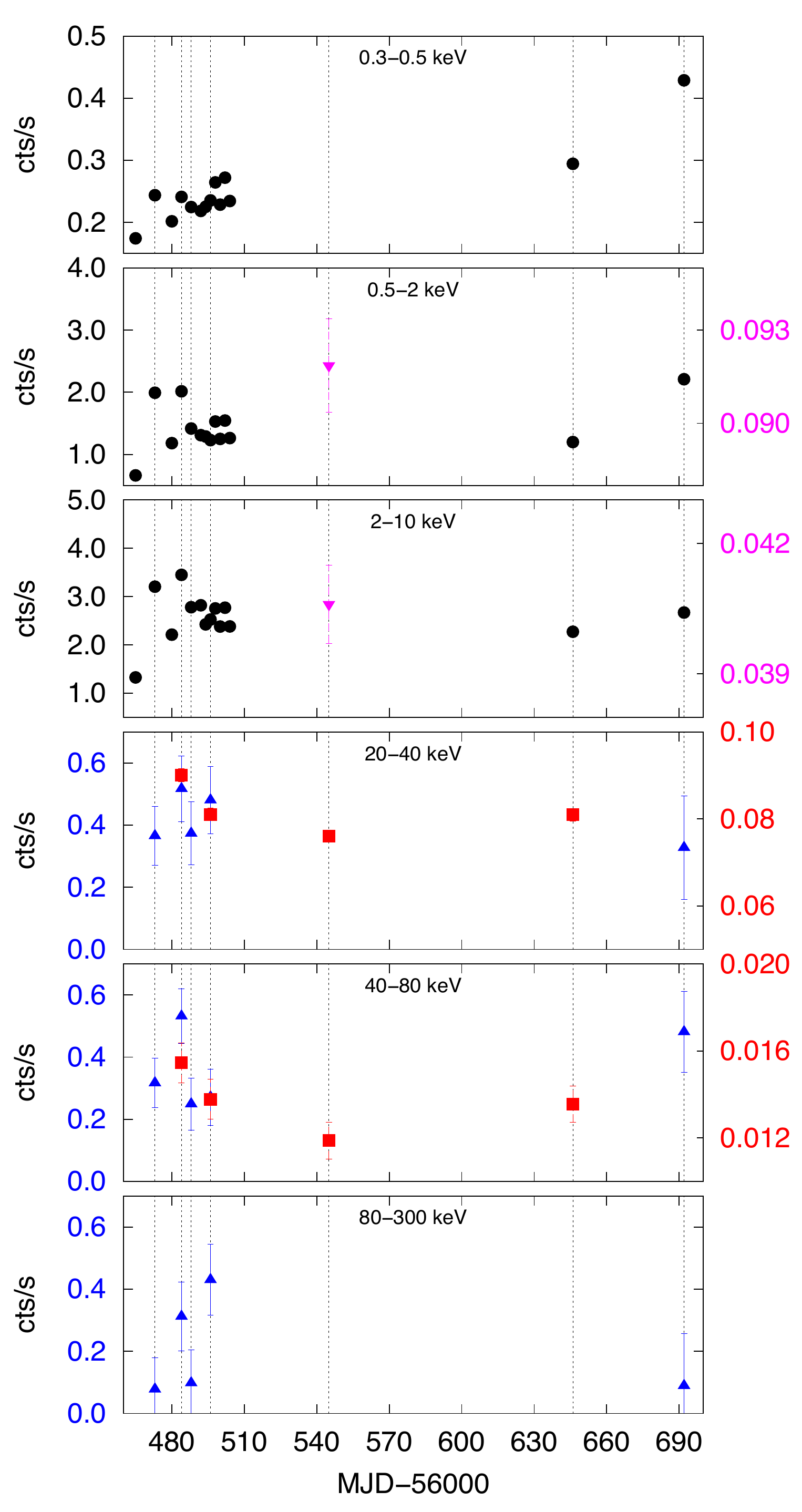}
\caption{Light curves in different energy ranges. On the x-axis, we report the last three digits of the Modified Julian Date. On the y-axis, on both left and right side, we report the count-rate. Black circles: \xmm (0.3-0.5 keV, 0.5-2 keV, 2-10 keV); magenta nablas: \chandra (0.5-2 keV, 2-10 keV); red squares: \nustar (20-40 keV, 40-80 keV); blue triangles: \integral (20-40 keV, 40-80 keV, 80-300 keV). The vertical dotted lines highlight the seven observations we analysed in this paper (i.e., those including \nustar and/or \integral data).
\label{lc}}
\end{figure}
\section{Observations \& Data Reduction}\label{sec:obs&data}
The multi-satellite campaign was led by \textit{XMM-Newton}, which observed NGC 5548 twelve times during 40 days in June and July 2013, plus once in December and once in February 2014. For the full description of the program, see \citetalias{kaastra2014science} and \citet{prova}. During the campaign, \nustar \citep{harrison2013nustar} and \integral \citep{winkler2003integral} regularly observed the source simultaneously with \textit{XMM-Newton}. A simultaneous \textit{Chandra/NuSTAR} observation was also performed in September 2013. The log of the data sets analysed in this paper is reported in Table \ref{obs}, with their observation ID, their start date, and net exposure time. The latest \xmm observation is dated February 4, and the corresponding spectrum has been analysed jointly with an average spectrum obtained from three \integral exposures obtained in January and February, which show no significant variations.\\ \\
\begin{table*}
\begin{center}
\caption{The logs of the simultaneous \xmmv \nustar and/or \integral observations of NGC 5548 during our campaign. \label{obs}}
\begin{tabular}{ c c c c c } 
\hline \rule{0pt}{2.5ex} Obs. &  Satellites & Obs. Id. & Start time (UTC)  & Net exp.\\ 
 & & & yyyy-mm-dd & (ks)  \\ \hline \rule{0pt}{2.5ex}
1 & \xmm & 0720110401&  2013-06-30  & 38\\ 
   & \integral & 10700010001 & 2013-06-29 &  62\\ \hline \rule{0pt}{2.5ex}
2 & \xmm & 0720110601 & 2013-07-11 &  37 \\
   & \nustar & 60002044002/3 & 2013-07-11  & 50\\
   & \integral & 10700010002 & 2013-07-11  &50\\ \hline \rule{0pt}{2.5ex}
3 & \xmm & 0720110701 &    2013-07-15  & 37\\
   & \integral & 10700010003 & 2013-07-15  & 50 \\ \hline \rule{0pt}{2.5ex}
4 & \xmm & 0720111101& 2013-07-23  & 38\\ 
   & \nustar &60002044005 & 2013-07-23  & 50 \\ 
   & \integral &10700010004 & 2013-07-23  & 52 \\ \hline \rule{0pt}{2.5ex}
5 & \chandra & 16314 & 2013-09-10  &120 \\ 
   & \nustar & 60002044006 & 2013-09-10  & 50\\ \hline \rule{0pt}{2.5ex}
6 & \xmm & 0720111501 & 2013-12-20  & 38 \\
   & \nustar & 60002044008 & 2013-12-20  & 50\\ \hline \rule{0pt}{2.5ex}
7 & \xmm & 0720111601&  2014-02-04  & 38 \\ 
   & \integral & 11200110001 & 2014-01-17  & 94\\
   && 11200110002 & 2014-01-22  & 40\\
   && 11200110003 & 2014-02-09  & 30\\ \hline
 \end{tabular}
\end{center}
\end{table*}
In all the \xmm observations, the EPIC instruments were operating in the Small-Window mode with the thin-filter applied. For our analysis, we used the EPIC-pn data only. The data were processed using the \xmm Science Analysis System (\textsc{sas} v13). The spectra were grouped such that each spectral bin contains at least 20 counts. Starting in 2013, a gain problem with the pn data appeared, causing a shift of about 50 eV near the Fe-K complex (see  the detailed discussion in Cappi et al., in prep.). In the present paper this shift was corrected for by leaving the redshift free to vary in our models. Moreover, this small offset yields poor fits near the energy of the gold M-edge of the telescope mirror, and for this reason the interval between 2.0 and 2.5 keV was omitted from the fit. However, this problem has no effect upon the broad-band modelling, in particular the Fe-K line strength is correct, thus allowing for the reflection component to be correctly modelled.\\ \\
The \nustar data were reduced with the standard pipeline (\textsc{nupipeline}) in the \nustar Data Analysis Software (\textsc{nustardas}, v1.3.1; part of the \textsc{heasoft} distribution as of version 6.14), using calibration files from \nustar {\sc caldb} v20130710. The raw event files were cleaned with the standard depth correction, which greatly reduces the internal background at high energies, and data from the passages of the satellite through the South Atlantic Anomaly were removed. 
Spectra and light curves were extracted from the cleaned event files using the standard tool \textsc{nuproducts} for each of the two hard X-ray telescopes aboard \textit{NuSTAR}, focal plane modules A and B (FPMA and FPMB); the spectra from these telescopes are analysed jointly in this work, but are not combined. Finally, the spectra were grouped such that each spectral bin contains at least 50 counts.\\ \\
\integral data were reduced with the Off-line Scientific Analysis software \textsc{osa} 10.0 \citep{courvoisier2003integral}. IBIS/ISGRI spectra have been extracted using the standard spectral extraction tool {\tt ibis\_science\_analysis} and combined together using the \textsc{osa} {\tt spe\_pick} tool to generate the total ISGRI spectrum, as well as the corresponding response and ancillary files.\\ \\
The \chandra HRC/LETG spectra (first order) of the 120 ks observation of September 10 were reduced with the package \textsc{ciao} v4.6 and reprocessed with the standard tool {\tt chandra\_repro}.\\ \\
Finally, we made use of archival {\it Beppo}SAX and {\it Suzaku} data to compare the results of the campaign with past X-ray observations of the source. We focused on a 1997 {\it Beppo}SAX long ($\sim$300 ks) observation and on seven {\it Suzaku} observations ($\sim$30 ks each) performed in 2007. For a description of the observations and data reduction procedure we refer to \cite{nicastro5548} for {\it Beppo}SAX, and \cite{5548suzaku}  for {\it Suzaku}. \\ \\
In Fig. \ref{lc}, we show the light curves of all the fourteen \xmm observations, the four \textit{NuSTAR} observations and the five \integral observations of the campaign. The variability shown by the light curves in the low-energy ranges is mainly due to obscuration, as shown in Sec. \ref{sec:analysis}. We will also show that the primary emission exhibits intrinsic variations as well, which explain the variability observed at  higher energies. We refer the reader to \citet{prova} for the full time-line of the campaign (see their Fig. 1). 

\section{Spectral analysis}\label{sec:analysis}
Spectral analysis and model fitting was carried out with the \textsc{xspec} 12.8 package \citep{arnaud1996}, using the $\chi^2$ minimisation technique throughout. In this paper, all errors are quoted at the 90\% confidence level.
\subsection{Spectral curvature above 10 keV}\label{subsec:curvature}
\begin{figure}
\includegraphics[width=\linewidth]{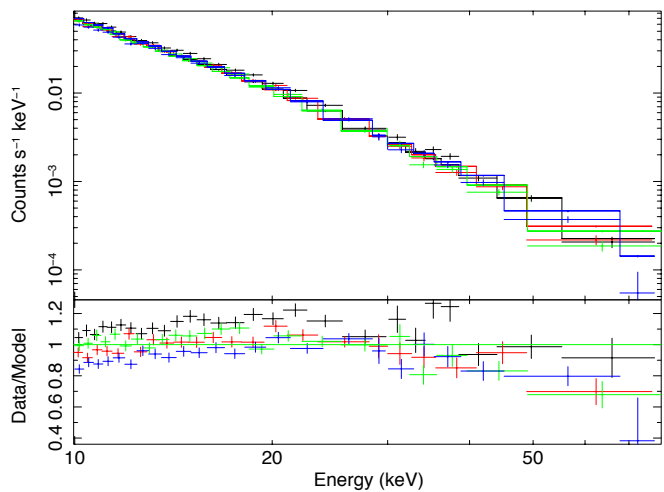}
\caption{Upper panel: the 10-79 keV spectra of the four \nustar observations. Lower panel: residuals when simultaneously fitting with a power law. For clarity, we used only FPMA data, which are rebinned in the plot.
\label{bat}}
\end{figure}

In Fig. \ref{bat} we plot for comparison the four {\it NuSTAR}/FPMA spectra above 10 keV, simultaneously fitted with a power law with tied parameters ($\chi^2$/dof = 1549/1178). The photon index obtained is $\Gamma = 1.58 \pm 0.02$. The plot shows that the spectral variability in the hard X-rays during the campaign is not dramatic, while it is prominent in the soft X-rays (\citetalias{kaastra2014science}; Cappi et al., in prep.; di Gesu et al., in prep.) where obscuration plays a major role. 
After getting a general idea about the power law parameters, we tested the presence of spectral curvature by individually fitting each of the four \nustar spectra above 10 keV, first with a simple power law and as a second step including an exponential cut-off. The addition of the cut-off results in an improvement in all fits ($\Delta \chi^2 \lesssim -30$), with a probability of chance improvement less than 10$^{-8}$ according to the $F$-test. This result motivates the use of a cut-off power law in the analysis of the data. However, as a next step we test whether the spectral curvature is partly due to a reflection component.
\subsection{The reflection component}\label{subsec:refl}
\begin{figure}
\includegraphics[width=\linewidth]{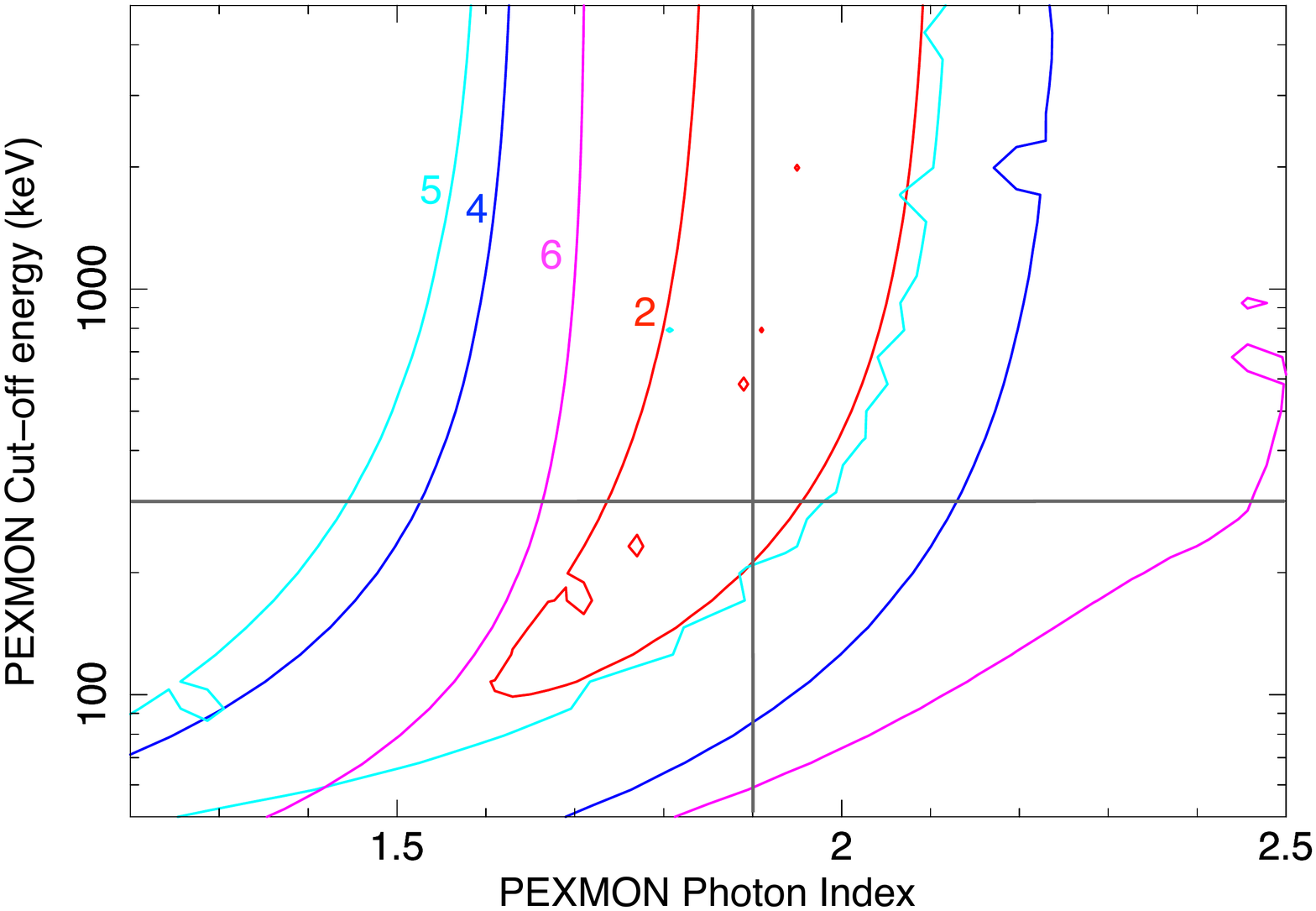}
\caption{\textsc{pexmon} cut-off energy vs. photon index 90\% confidence level contours for the 4 \nustar observations, fitted above 5 keV. The grey solid lines indicate the values we chose to fix in the later broad-band analysis (1.9 for the photon index, 300 keV for the cut-off energy).
\label{plotcont1}}
\end{figure}

\begin{figure}
\includegraphics[width=\linewidth]{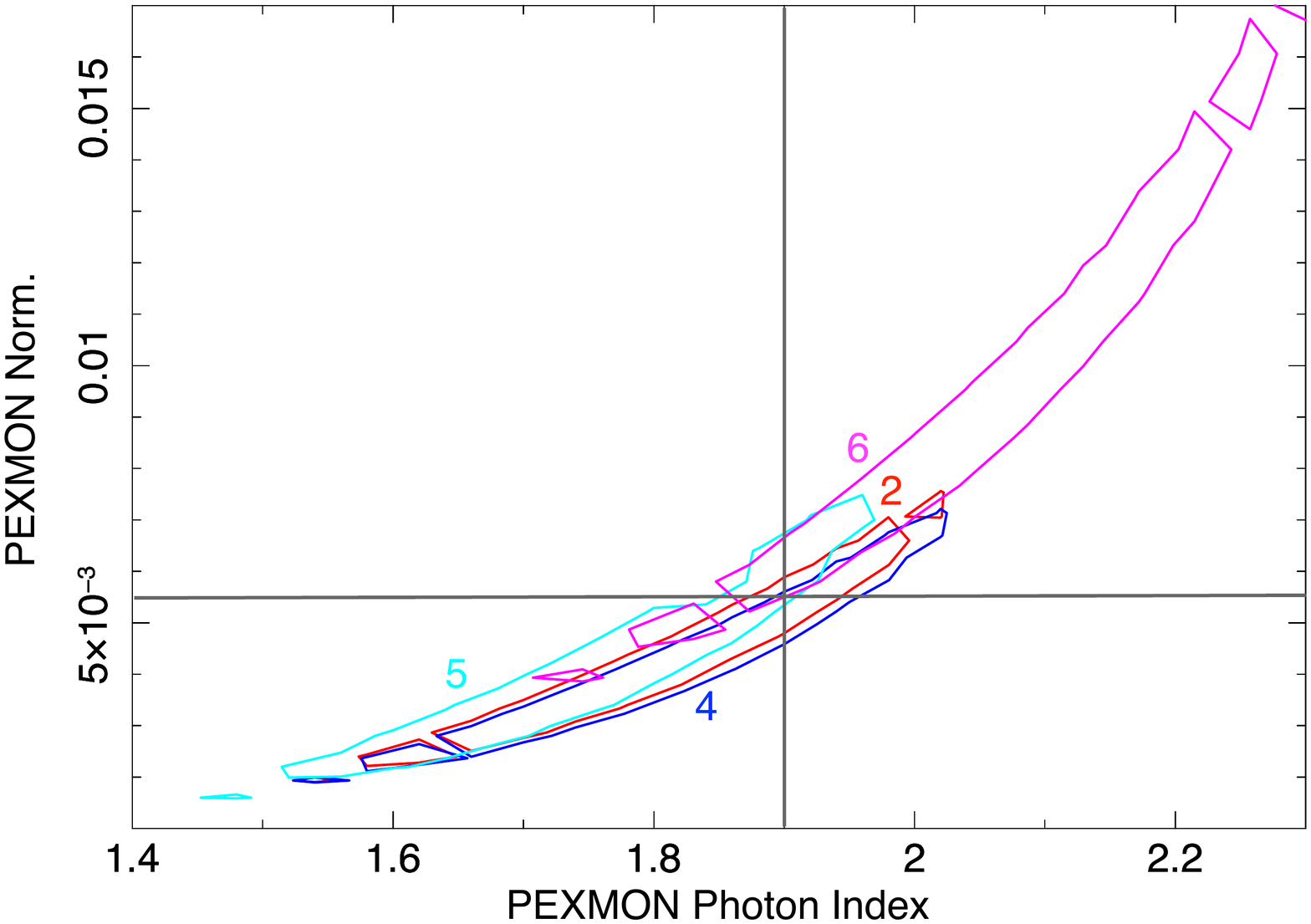}
\caption{\textsc{pexmon} normalization vs. photon index 90\% confidence level contours for the 4 \nustar observations, fitted above 5 keV. The grey solid lines indicate the values we chose to fix in the later broad-band analysis (1.9 for the photon index, $5.5 \times 10^{-3}$ for the normalization).
\label{plotcont2}}
\end{figure}

To test for the presence of a reflection component, we fitted the data above 5 keV. By choosing 5 keV as a lower energy limit, we avoid the effects due to the strong absorption in the soft band. Good statistics are needed, so we only consider the four observations including \nustar data (obs. 2, 4, 5 and 6 of Table \ref{obs}).\\ We fitted the \textit{XMM-Newton}/pn, \nustar and \integral data simultaneously, leaving the cross-calibration constants free to vary. The normalization factors with respect to pn were always between 1 and 1.1. In particular, the spectral agreement between \xmm and \nustar is generally good in their common 3-10 keV bandpass \citep{walton13, walton14}. 
We used a simple model, consisting of a cut-off power law  modified by absorption, plus reflection. We included a neutral absorber with covering fraction fixed to 1, while the column density was free to vary. This is clearly an arbitrary choice, but the way we model the absorption, with a lower energy limit of 5 keV, does not affect the results that follow.\\ \\
Following the analysis done by Cappi et al. (in prep.) and motivated by their detection of a neutral, narrow and constant Fe K$\alpha$ line, we used the \textsc{pexmon} model \citep{nandra2007pexmon} in \textsc{xspec} for the reflection component. \textsc{pexmon} combines a neutral Compton reflection, as described by the \textsc{pexrav} model \citep{pexrav} in \textsc{xspec}, with Fe K$\alpha$, Fe K$\beta$, Ni K$\alpha$ lines and the Fe K$\alpha$ Compton shoulder, generated self-consistently by Monte-Carlo simulations \citep{georgefabian1991}. We fixed the inclination angle of the reflector to 30 deg, appropriate for a type 1 source \citep[e.g.,][]{nandra1997} and consistent with past analyses of the Fe K lines in NGC 5548 \citep{yaqoob5548,5548suzaku}, but also with the inclination estimated for the broad-line region \citep{pancoast2013mbh5548} and for the disk-corona system \citep{pop5548}. The photon index, cut-off energy and normalization of the incident radiation were free to vary and not tied to those of the primary power law. This choice is justified \textit{a priori} by the observed constancy of the Fe K$\alpha$ line, and \textit{a posteriori} by our constraints on the reflection component, which is consistent with being constant during the campaign. In this case, the distance of the reflecting material from the X-ray emitting region should be at least a few light months, i.e. a few $\times$ $10^4$ gravitational radii, as we will discuss in Sect. \ref{sec:discussion}. \\ \\
The 90\% contour plots of cut-off energy and normalization vs. photon index of \textsc{pexmon} are plotted in Figs. \ref{plotcont1} and \ref{plotcont2}. The photon index and cut-off energy of the reflected continuum are not well constrained, however the contours are mutually consistent. The normalization vs. photon index contours are very similar for the four observations, with indication of a slightly larger normalization for obs. 6; nevertheless they are also marginally consistent with each other. These results are consistent with the reflected emission coming from remote material, and thus the parameters being constant during the campaign. For this reason, we chose to freeze hereafter the parameters of the reflection component to their weighted average derived from our analysis. We thus fixed the \textsc{pexmon} photon index to 1.9, cut-off energy to 300 keV and normalization to $5.5 \times 10^{-3}$ (\textsc{xspec} units). The photon index of 1.9 is relatively high compared to the average value for NGC 5548, which is $\sim 1.7$ \citep[see, e.g.,][]{sobolewska&papadakis} but has indeed been observed in the past \citep[][]{nandra19915548,reynolds1997,chiang20005548}. More importantly, the \textit{Swift} monitoring discussed in Mehdipour et al. (in prep.) shows that the photon index reached 1.8-1.9 during most of 2012. Finally, from the \textsc{pexmon} parameters we estimate the average reflection fraction $f$ as $0.3 \pm 0.1$, calculated as the ratio between the 20-40 keV flux illuminating the reflecting material and the 20-40 keV flux of the primary cut-off power law\footnote{This definition is slightly different from the traditional $R=\Omega/2\pi$, where $\Omega$ is the solid
angle subtended by the reflector \citep{pexrav}.}. For a distant reflector, this ratio might not have a clear link with the covering factor as seen from the X-ray source \citep{malzac&petrucci2002}, but this is discussed in more detail in Sec. \ref{sec:discussion}. 
\subsection{Broad-band spectral analysis}\label{subsec:broad-band}
Having obtained constraints on the parameters of the reflection component using the data above 5 keV, we now proceed to fit the total X-ray energy range available. We first use a phenomenological model, which is described below. Given the different satellites used in the seven observations, the energy range is limited to 0.5-79 keV for obs. 5 (\textit{Chandra/NuSTAR}) and 0.3-79 keV for obs. 6 (\textit{XMM-Newton/NuSTAR}). For all the other observations, the energy range is 0.3-400 keV, thanks to the addition of the \integral data.
\subsubsection{The model components}\label{subsec:components}
Following Sect. \ref{subsec:curvature} and \ref{subsec:refl}, we modelled the continuum with a cut-off power law and the reflected component with \textsc{pexmon}, whose parameters were fixed.  We describe in the following the other spectral components.\vspace{-6pt}
\paragraph{Obscuration:} 
The model presented by \citetalias{kaastra2014science} includes two ionized, partially covering absorbers. Both were found to have a low degree of ionization, namely $\log \xi \leq -4$ and $\log \xi \simeq -1.2$, such that at the EPIC-pn energy resolution they are almost equivalent to neutral absorbers. In particular, the absorber with $\log \xi = -1.2$ was found to have a column density of $1.2 \times 10^{22}$ cm$^{-2}$ \citepalias{kaastra2014science} and such an absorber is indistinguishable, using pn data only, from a neutral one with a column density of $1.1 \times 10^{22}$ cm$^{-2}$. Since a detailed modelling of the obscuration is out of the scope of the present paper, we modelled the two absorbing components with \textsc{pcfabs} in \textsc{xspec}, which describes partial covering absorption by a neutral medium. Both the column density and the covering fraction of the two components were free to vary between different observations.\vspace{-6pt}
\paragraph{The warm absorber:} The model by \citetalias{kaastra2014science} includes six warm absorber components, with different column densities ($N_H\sim 10^{20}$-$10^{21}$ cm$^{-2}$), ionization parameters ($\log \xi \sim$ 1-3), and outflow velocities ($v \sim$ 250-1000 km/s). We produced a table, with fixed parameters, from the best-fit model of the warm absorber derived by \citetalias{kaastra2014science} and included it in \textsc{xspec}. For a comprehensive analysis of the UV properties of the warm absorber, see 
\citet{arav}.\vspace{-6pt}
\paragraph{The soft X-ray excess:} The physical origin of this component is still under debate, and it can be modelled with several different spectral models. The most common interpretations rely on relativistically blurred reflection \citep[see, e.g.,][]{crummy2006,cerruti2011,walton2013} or Comptonization by a warm medium \citep[see, e.g.,][]{pop2013mrk509,rozenn2014mrk509SE,matt2014ark120}. Detailed modelling of the soft X-ray excess is beyond the scope of this paper, we thus followed the analysis done by Mehdipour et al. (in prep.) and assumed a thermal Comptonization component modelled by \textsc{comptt} \citep{titarchuk1994} in \textsc{xspec}. The parameters of \textsc{comptt} were frozen, for each individual observation, to the values reported by Mehdipour et al. (in prep.), to which we refer the reader for details. We note that Mehdipour et al. (in prep.) found a significant flux variability of the soft excess between different observations, which we take into account here, but no apparent spectral shape variability.\vspace{-6pt}
\paragraph{Narrow emission lines:} A number of emission lines, such as [O VII] at 0.56 keV and [C V] at 0.3 keV, were detected in the \xmm RGS spectra, as shown by \citetalias{kaastra2014science}. Again, we produced a table from the best-fit model and included it in \textsc{xspec}. These lines arguably originate from the Narrow-Line Region, therefore we do not expect line flux variations during the campaign \citep{peterson2013}.
\subsubsection{Best-fit results}
\begin{figure}
\includegraphics[width=\linewidth]{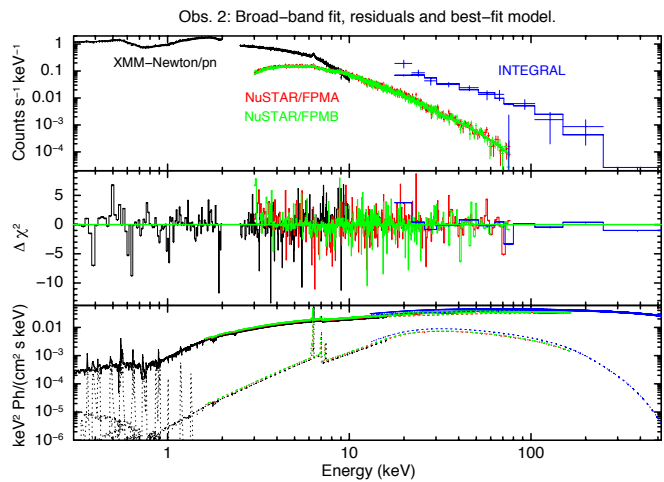}
\caption{Data and best-fit model for obs. 2 ({\it XMM-Newton/NuSTAR/INTEGRAL}, see Table \ref{obs}). Upper panel: broad-band data and folded model. Middle panel: contribution to $\chi^2$. Lower panel: best-fit model $E^2 f(E)$ with the plot of the various additive components, namely the cut-off power law, the reflection component, the soft excess and the narrow lines.
\label{ldata}}
\end{figure}

\begin{figure}
\includegraphics[width=\linewidth]{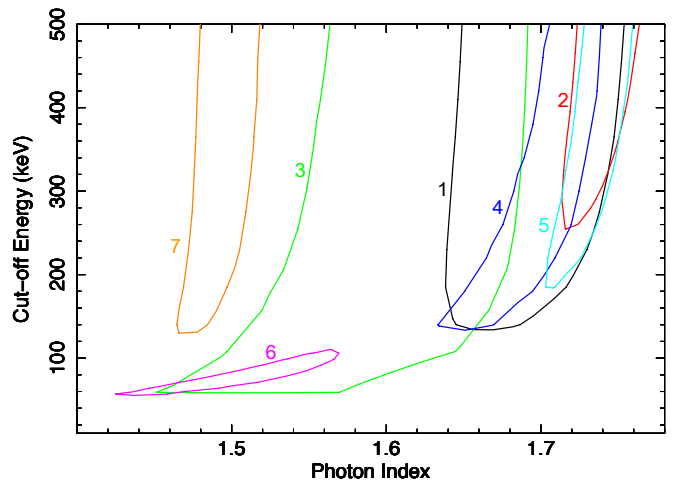}
\caption{Cut-off energy vs. primary photon index 90\% confidence level contours for obs. 1 (black), obs. 2 (red), obs. 3 (green), obs. 4 (blue), obs. 5 (cyan), obs. 6 (magenta), obs. 7 (orange).
\label{cont1}}
\end{figure}

\begin{figure}
\includegraphics[width=\linewidth]{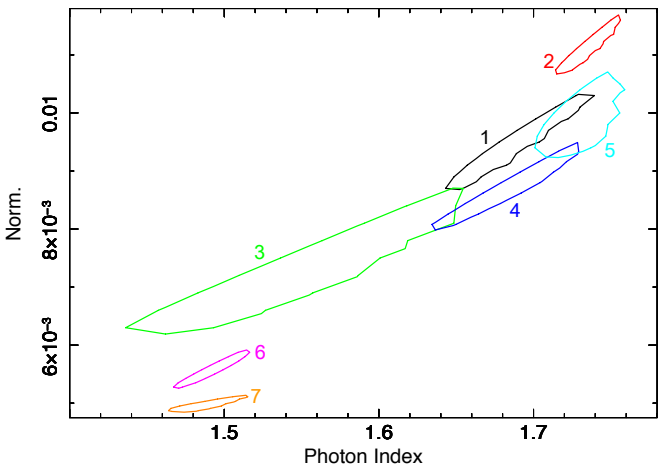}
\caption{Normalization vs. primary photon index 90\% confidence level contours for obs. 1 (black), obs. 2 (red), obs. 3 (green), obs. 4 (blue), obs. 5 (cyan), obs. 6 (magenta), obs. 7 (orange).
\label{cont2}}
\end{figure}

\begin{figure}
\includegraphics[width=\linewidth]{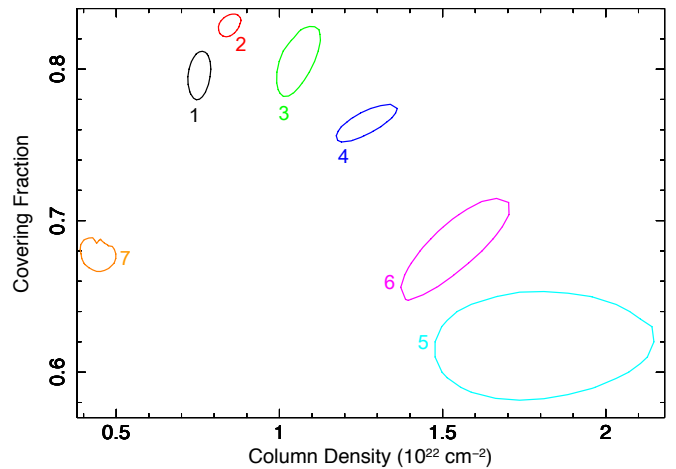}
\caption{Obscurer component 1: $N_H - C_F$ 90\% confidence level contours for obs. 1 (black), obs. 2 (red), obs. 3 (green), obs. 4 (blue), obs. 5 (cyan), obs. 6 (magenta), obs. 7 (orange).
\label{nh1}}
\end{figure}

\begin{figure}
\includegraphics[width=\linewidth]{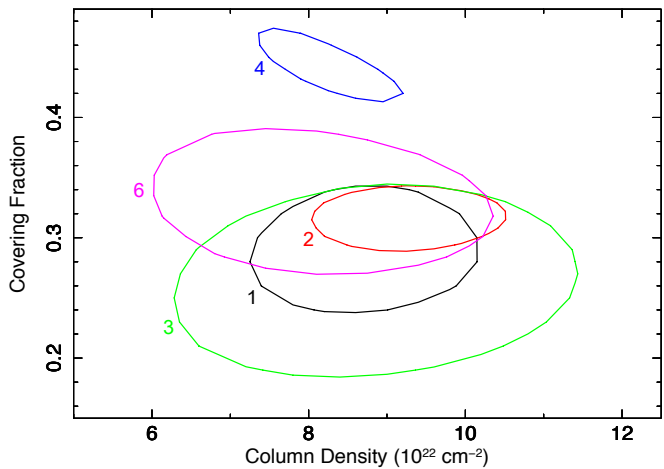}
\caption{Obscurer component 2: $N_H - C_F$ 90\% confidence level contours for obs. 1 (black), obs. 2 (red), obs. 3 (green), obs. 4 (blue), obs. 6 (magenta). This obscurer was not found in obs. 5 and 7. 
\label{nh2}}
\end{figure}

\begin{figure}
\includegraphics[width=\linewidth]{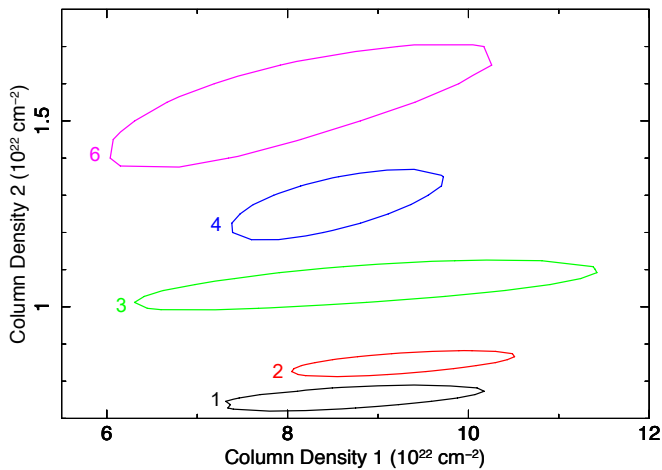}
\caption{$N_H - N_H$ 90\% confidence level contours for obs. 1 (black), obs. 2 (red), obs. 3 (green), obs. 4 (blue), obs. 6 (magenta).
\label{nh3}}
\end{figure}

The best-fit parameters are reported in Table \ref{fitbroad}, while in Fig. \ref{ldata} we show the data, residuals and best-fit model for obs. 2 (\textit{XMM-Newton/NuSTAR/INTEGRAL}). 
The broad-band fit allows us to constrain the spectral shape more reliably with respect to the fit above 5 keV. In Fig. \ref{cont1} and \ref{cont2}, we show the contour plots of the high-energy cut-off and normalization versus photon index, both for the primary continuum. 
The contours show significant variability of the photon index from $\sim1.5$ (obs. 6, 7) to $\sim1.7$ (obs. 1, 2, 4, 5). It should be noted that the photon index given in Kaastra et al. (2014), i.e. $\Gamma = 1.566 \pm 0.009$, is a mean value which comes from the analysis of the co-added \xmm and \nustar spectra of summer 2013. By including the individual \nustar and \integral high-energy spectra, we could constrain the photon index for each observation. Furthermore, the cut-off energy shows significant variability as well. We derive a lower limit for the cut-off energy in all the observations, and a measure of $70^{+40}_{-10}$ keV for obs. 6. This measure is inconsistent, at 90\% confidence level, with the lower limits determined from the other observations, with the exception of obs. 3. Finally, we note that the constraints we derived on the cut-off energy are mostly compatible with the average value estimated in sec. \ref{subsec:refl} for the reflection component alone, i.e. 300 keV.\\ \\
The broad-band fit helps constrain the obscurer components as well. In Fig. \ref{nh1} and \ref{nh2}, we plot the contours of the covering fraction versus the column density of the two neutral absorbers, while in Fig. \ref{nh3} we plot the contours of the two column densities versus each other. These contours show that absorption variability is statistically significant for both the low and high-$N_H$ absorbers. This can be seen, for example, by comparing obs. 2 (red contours) with obs. 4 (blue contours). Since the time interval between obs. 2 and 4 is $\sim10$ days, we can conclude that the variability is significant on this time scale. In two cases, namely obs. 5 and 7, only one, low-$N_H$ absorber is needed, and adding the high-$N_H$ component does not improve the fit. For obs. 5 this can be explained by the lower sensitivity of \chandra HRC/LETG with respect to \xmm EPIC-pn, which results in a lower {\it S/N} in the soft band. For obs. 7, which is the last one and dated February 4, the absence of the high-$N_H$ component could simply mean that the source started changing back to the more usual, unobscured condition. However, for the model described above, we only obtain a relatively poor fit ($\chi^2$/dof = 651/424), with strong, positive residuals around 0.5 keV, and negative residuals around 0.8 keV. We stress that the models we used for the warm absorber and the narrow emission lines were obtained from the analysis of pn and RGS data \citepalias{kaastra2014science}, while our analysis is based on pn data alone. We may thus attribute such residuals in obs. 7 to a combination of the previously mentioned gain problem of the pn data with cross-calibration issues between the pn and RGS instruments \citep{xmm_calibration}. We obtained a much better fit ($\Delta \chi^2 \simeq -200$) by adding a narrow Gaussian line with \textsc{gauss} in \textsc{xspec}, and an absorption edge with \textsc{edge}, without affecting the parameters of the primary cut-off power law. The emission line was found at $E=0.54\pm0.01$ keV, with an upper limit on the equivalent width of 20 eV. The absorption edge was found at $E=0.80\pm0.01$ keV with $\tau=0.28\pm0.06$. 
\begin{table*}
\begin{center}
\caption{Best-fit parameters of the primary cut-off power law and the two absorbers (1) and (2), fitting the data down to 0.3 keV. In the case of obs. 7, the model included a Gaussian line and an absorption edge (see text). \newline $E_c$ is in keV, $F_{1\, \textrm{keV}}$ is in units of $10^{-2}$\,ph/keV cm$^2$ s and $N_H$ is in units of $10^{22}$ cm$^{-2}$. \newline $\dag$ The second absorber was not needed in obs. 5 (\chandra plus \textit{NuSTAR}) and obs. 7. In these two cases, fixing $N_H^{(2)}$ to 8.5 resulted in $C_F^{(2)}=0$.\label{fitbroad}}
\begin{tabular}{ c c c c c c c c c} 
\hline \rule{0pt}{3ex}  Obs. & $\Gamma$ & $E_c$ & $F_{1\, \textrm{keV}}$ & $N_H^{(1)}$  & $C_F^{(1)}$  & $N_H^{(2)}$ & $C_F^{(2)}$  & $\chi^2$/dof\\ \hline \rule{0pt}{3ex} 
1 & $1.70^{+0.04}_{-0.05}$ & >120& $0.96\pm 0.09$ & $0.75\pm 0.03$ & $0.797^{+0.006}_{-0.007}$ & $9 \pm 1$ & $0.29^{+0.05}_{-0.06}$ & 538/424\\  \rule{0pt}{3ex}
2 & $1.74^{+0.01}_{-0.04}$ & >250& $1.13^{+0.03}_{-0.07}$ & $0.85\pm 0.02$ & $0.829^{+0.004}_{-0.003}$ & $9.2^{+0.8}_{-0.9}$ & $0.32^{+0.02}_{-0.03}$ & 1587/1393\\  \rule{0pt}{3ex}
3 & $1.6^{+0.1}_{-0.2}$ & >50 & $0.7^{+0.2}_{-0.1}$ & $1.06^{+0.06}_{-0.10}$ & $0.81^{+0.01}_{-0.02}$ & $9^{+3}_{-2}$ & $0.27^{+0.06}_{-0.10}$  & 508/420\\  \rule{0pt}{3ex}
4 & $1.68\pm 0.05 $ & >120 & $0.87^{+0.07}_{-0.06}$ & $1.27\pm 0.06$ & $0.77\pm 0.01$  & $8.5^{+0.8}_{-0.7}$ & $0.46\pm 0.03$ & 1346/1329\\  \rule{0pt}{3ex}
5 & $1.74^{+0.01}_{-0.03} $ & >180 & $1.00\pm 0.05$ & $1.8\pm 0.02$ & $0.62\pm 0.03$ &  $\dag$ & $\dag$ &1113/1094\\ \rule{0pt}{3ex}
6 & $1.49\pm 0.05 $ & $70^{+40}_{-10}$ & $0.56\pm 0.05$ & $1.5^{+0.2}_{-0.1}$ & $0.68\pm 0.04$ &  $8 \pm 2 $& $0.33\pm 0.06$ & 1484/1308\\ \rule{0pt}{3ex}
7 & $1.49\pm 0.02 $ & >120 & $0.50\pm 0.01$ & $0.44^{+0.05}_{-0.04}$ & $0.678\pm 0.007$ & $\dag$ & $\dag$ & 452/420\\ \hline
 \end{tabular}
\end{center}
\end{table*}
\subsection{Test of Comptonization model}\label{subsec:compt}
\begin{figure}
\includegraphics[width=\linewidth]{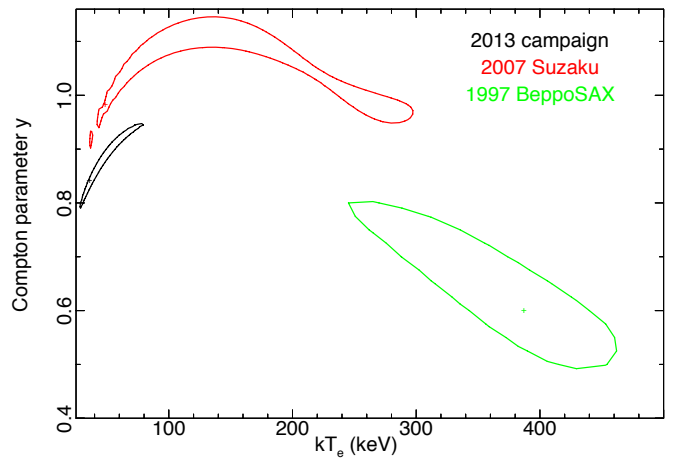}
\caption{$kT_e - y$ 90\% confidence level contours for the spectra fitted with \textsc{compps}, using spherical geometry. Black: average spectrum of the 2013 campaign; red: average spectrum of the seven \textit{Suzaku} observations taken in 2007; green: \textit{Beppo}SAX observation taken in 1997.
\label{kt_tau}}
\end{figure}

The last step of our analysis consists of fitting the average spectrum with a more realistic model, where the primary continuum is produced self-consistently via Comptonization of soft photons emitted by the disk. A more detailed analysis of the entire campaign with realistic Comptonization models is deferred to a future work.\\ \\
First, we produced an average spectrum merging respectively the \xmmv \nustar and \integral spectra used in this paper (Table \ref{obs}) with the \textsc{mathpha} tool of \textsc{heasoft}. We then proceeded to fit this average spectrum over the range 0.3-400 keV, at first with the phenomenological model we have discussed above (Sect. \ref{subsec:components}). The resulting fit had $\chi^2$/dof = 3314/2631, with strong, positive residuals around 0.5 keV, and negative residuals around 0.6 keV (analogously to the already discussed obs. 7). We thus included two narrow Gaussian lines, with $E=0.53\pm0.06$ keV (in emission, equivalent width $<$ 50 eV) and $E=0.60\pm0.02$ keV (in absorption, equivalent width $<$ 20 eV). Again, we note that such residuals are most likely unphysical and due to cross-calibration between pn and RGS \citep{xmm_calibration}. However, the addition of these lines, while improving the fit, does not affect the best-fit parameters of the cut-off power law, namely $\Gamma=1.67\pm0.02$ and $E_c > 250$ keV, with $\chi^2$/dof=3080/2627 and no prominent residual features. The relatively large $\chi^2$ is most likely an effect of averaging variable spectra (in particular in the soft band, due to absorption variability), and of the not perfect cross-calibration between \xmm and {\it NuSTAR}, which especially appears in the co-added spectrum. We also note that \xmm and \nustar do not cover strictly simultaneous times, since the exposure time of the two satellites is slightly different (see Table \ref{obs}). \\ \\
We then replaced the cut-off power law with the Comptonization model \textsc{compps} \citep{compps} in \textsc{xspec}. \textsc{compps} models the thermal Comptonization emission of a plasma cooled by soft photons with a disk black-body distribution. We chose a spherical geometry (parameter \textsc{geom}=-4 in \textsc{compps}) for the hot plasma and a temperature at the inner disk radius of 10 eV, fitting for the electron temperature $kT_e$ and Compton  parameter $y = 4 \tau (kT_e/m_ec^2)$. The best-fit parameters are reported in Table \ref{compps} and the $kT_e - y$ contour plot is shown in Fig. \ref{kt_tau} (black contour). From the best-fit temperature $kT_e = 40$ keV and parameter $y=0.85$, we derive an optical depth $\tau = 2.7^{+0.7}_{-1.2}$, where we estimate the errors from the $kT_e - y$ contour.
\begin{table*}
\begin{center}
\caption{Best-fit parameters of the average data set, using the realistic Comptonization model \textsc{compps} for the continuum. The physical units are the same as in Table \ref{fitbroad}. \newline The normalization is defined as $K=R_{\textrm{km}}/D_{\textrm{10 kpc}}$ where $R_{\textrm{km}}$ is the source radius in km and $D_{\textrm{10 kpc}}$ the distance in units of 10 kpc. \label{compps}}
\begin{tabular}{ c c c c c c c c c} 
\hline \rule{0pt}{3ex} $kT_e$ (keV) & $y$ & $K$ ($\times 10^{7}$) & $N_H^{(1)}$ & $C_F^{(1)}$  & $N_H^{(2)}$ & $C_F^{(2)}$  & $\chi^2$/dof\\ \hline \rule{0pt}{3ex}
$40^{+40}_{-10}$ & $0.85^{+0.10}_{-0.5}$ & $6.0^{+0.4}_{-0.3}$ & $0.99 \pm 0.02$ & $0.766 \pm 0.003$  &  $9.6 \pm 0.5$ & $0.36 \pm 0.01$ & 3079/2627\\ \hline
\end{tabular}
\end{center}
\end{table*}
\section{Discussion and conclusions}\label{sec:discussion}
The broad-band coverage of the \textit{XMM-Newton, NuSTAR} and \textit{INTEGRAL} data allowed us to study the different components of the high-energy spectrum of NGC 5548 in detail.\\ \\
First, we have been able to constrain the reflected component, which is consistent with being constant on a few months time scale, while the primary emission exhibits significant variability on a few days time scale. For example, the flux in the 2-10 keV band exhibits variations up to a factor of two over about 5 days during the campaign, with an outburst of several days duration in August-September \citep[][see their Fig. 2]{prova}. The lack of variability of the reflection component indicates that reflection is produced by cold and distant matter, lying at least a few light months away from the central X-ray source. This is consistent with the analysis of the Fe K$\alpha$ line by Cappi et al. (in prep.) and with the absence of a significant contribution from relativistic reflection in this source \citep[see, e.g.,][]{brenneman20125548}. The lower limit of the reflector distance is consistent with this material being in the putative dusty torus, as the dust sublimation radius is estimated to be $\sim 5$ light months \citep{crenshaw2003wa5548}. Furthermore, our results can be compared with past observations of the source. For this we focus on the 1997 \textit{Beppo}SAX observation of NGC 5548 \citep{nicastro5548,pop5548} and the average of seven \textit{Suzaku} observations conducted in 2007 \citep{5548suzaku}. In both cases, we re-fitted the archival data above 5 keV following the procedure described in Sect. \ref{subsec:refl}. For the reflection component, we find that the spectral shape is always consistent with our best-fit parameters, i.e. a photon index of 1.9 and cut-off energy of 300 keV, while the reflected flux shows significant variations of about 50\%. The primary flux (e.g., in the 2-10 keV band) shows similar variations, and we obtain a reflection fraction $f$ consistent with that found during the 2013 campaign, i.e.  $f \sim 0.3$, again calculated as the ratio between the illuminating and primary flux in the 20-40 keV range (see end of Sect. \ref{subsec:refl}). Since $f$ is consistent with being constant over a few years, the reflecting material should lie within a few light years from the primary X-ray source. In this case, the reflection fraction is indicative of the covering fraction of the reflector as seen from the X-ray source. From the definition of $f$, it follows that $f=\Omega / (4 \pi - \Omega)$. According to our results, then, the solid angle covered by the reflecting material can be estimated as $\Omega \simeq \pi$, consistently with previous findings \citep[e.g.,][]{nandra&pounds1994,pounds2003}. \\ \\
Then, we have been able to constrain the primary continuum and the effects of absorption. The primary continuum is well fitted by a power law with a photon index varying between $\sim$1.5 and $\sim$1.7, with at least a lower limit on the presence of a high-energy cut-off ($E_c>50$ keV), which is usually considered a manifestation of Comptonization and is expected to be linked to the coronal temperature. The approximate relation between the temperature $kT_e$ of an optically thick Comptonizing corona and the cut-off energy $E_c$ is $kT_e = E_c / 3$  \citep{poptestingcompt2001}. However, it should be noted that the cut-off of realistic Comptonization models is very different from an exponential cut-off. Moreover, the estimate of the temperature and of the optical depth also depends on the geometry assumed.
In one case (obs. 6) we have obtained a measure of $70^{+40}_{-10}$ keV for the cut-off energy, which provides an estimate of $23^{+14}_{-3}$ keV for the coronal temperature, with the caveats we just mentioned. Moreover, our results suggest a statistically significant variation of both the photon index and the cut-off energy. For example, in the case of obs. 5 we find a lower limit of 180 keV for the cut-off energy. We can thus estimate the coronal temperature to be $> 60$ keV for obs. 5 and $< 37$ keV for obs. 6, both at 90\% confidence level. Such a variation may be due to different reasons, like a variation of the disk-corona geometry. The coronal optical depth and temperature should follow a univocal relationship for a given heating/cooling ratio, i.e. the ratio of the power dissipated in the corona to the intercepted soft luminosity which cools the electrons via Comptonization \citep{haardt&maraschi1991,hmg1994,hmg1997}. If the coronal optical depth changes, due for instance to a change in the accretion rate, the temperature will change accordingly, to keep the heating/cooling ratio constant. But the heating/cooling ratio can also change, responding to a variation in the disk-corona geometry, like a variation of the inner disk radius or a velocity variation of a potentially outflowing corona \citep{malzac2001}. Then, the coronal optical depth and temperature will change in a non-trivial way to be consistent with the new heating/cooling ratio. Owing to these complexities, fitting the data with a phenomenological cut-off power law does not allow for a unique interpretation. However, our findings motivate us to apply realistic Comptonization models to constrain the coronal parameters. \\ \\ 
Concerning absorption, we needed to add two different partially covering components (only one in two cases out of seven). The obscurer component 1 has $N_H \sim 1 \times 10^{22}$ cm$^{-2}$ and $C_F \sim 0.8$, while the obscurer component 2 has $N_H \sim 9 \times 10^{22}$ cm$^{-2}$ and $C_F \sim 0.3$, consistently with the estimates of \citetalias{kaastra2014science}. Both show significant variability on a time-scale of a few days, indicating that the obscuring material varies along the line of sight. This relatively heavy obscuration was unexpected, since NGC 5548 is historically thought to be an archetypal unobscured Seyfert 1. The discussion reported in \citetalias{kaastra2014science} ascribes this obscuration to a fast and massive outflow which may be interpreted as a disk wind. \\ \\
Finally, we have made a first step towards a more physical description of the primary emission by replacing the simple cut-off power law with a thermal Comptonization model to fit the average spectrum. According to our analysis, the X-ray emitting corona has a mean temperature of $40^{+40}_{-10}$ keV and an optical depth of $2.7^{+0.7}_{-1.2}$, assuming a spherical geometry, during this campaign. We compared this result with past observations of NGC~5548 by re-fitting the 1997 \textit{Beppo}SAX spectrum and the 2007 {\it Suzaku} average spectrum. We used the \textsc{compps} model with spherical geometry (\textsc{geom}=-4) like we did in Sect. \ref{subsec:compt}. The $kT_e-y$ contours are shown in fig. \ref{kt_tau}. We find that the temperature and the optical depth are not consistent, at 90\% confidence level, between the different observations. According to this test, the average values of the coronal temperature and optical depth show a significant long-term variability, namely over 10-15 years, with a clear decrease of the coronal temperature from about 300-400 keV in 1997 down to 40-50 keV in 2013. The optical depth, on the other hand, increases from about 0.2-0.3 up to 2-3. Although the values quoted are specific to the assumed spherical geometry for the corona, the changes in temperature and optical depth are significant if one were to assume a different geometry for the corona. The corresponding variation of the Compton parameter $y$ suggests a variation, of uncertain origin, of the coronal heating/cooling ratio between these different observations.
The coronal parameters may be variable on shorter time scales as well, as suggested by the detected variations of the cut-off energy. \\ \\
Comparing with the 1997 \textit{Beppo}SAX observation and the 2007 \textit{Suzaku} observations, the 2013 campaign provided us a broader energy range which, together with the high signal-to-noise ratio spectra, allows for much better constraints on the coronal parameters (as shown by the contours in Fig. \ref{kt_tau}). However, for the optimal study of Comptonization, UV spectra may give a significant contribution. For example, \cite{pop2013mrk509} have been able to fit the UV and soft X-ray spectra of Mrk 509 with a Comptonization component interpreted as a "warm" corona above the disk. If this scenario is also valid for NGC 5548, we expect a correlation between the UV and soft X-ray emission. This correlation can only be detected after the variability of the obscurer is properly removed. The analysis of the UV and soft X-ray emission by Mehdipour et al. (in prep.) indeed suggests that such a correlation is present, and that it may be intrinsically similar to that of Mrk 509. This point, and a thorough test of realistic Comptonization models, will be the subject of a forthcoming paper.  

\section*{Acknowledgements}
We are grateful to the anonymous referee for his/her helpful comments, which have improved the manuscript.\\
This work is based on observations obtained with: the \textit{NuSTAR} mission, a project led by the California Institute of Technology, managed by the Jet Propulsion Laboratory and funded by NASA; \textit{INTEGRAL}, an ESA project with instrument and science data center funded by ESA member states (especially the PI countries: Denmark, France, Germany, Italy, Switzer- land, Spain), Czech Republic, and Poland and with the participation of Russia and the USA; \textit{XMM- Newton}, an ESA science mission with instruments and contributions directly funded by ESA Member States and the USA (NASA). The data used in this research are stored in the public archives of the international space observatories involved. This research has made use of data, software and/or web tools obtained from NASA's High Energy Astrophysics Science Archive Research Center (HEASARC), a service of Goddard Space Flight Center and the Smithsonian Astrophysical Observatory. 
FU, POP, GM, SB 
acknowledge support from PICS CNRS/INAF. FU, POP acknowledge support from CNES. FU acknowledges support from Universit\'e Franco-Italienne (Vinci PhD fellowship). FU, GM acknowledges financial support from the Italian Space Agency under grant ASI/INAF I/037/12/0-011/13. GAK was supported by NASA through grants for HST program number 13184 from the Space Telescope Science Institute, which is operated by the Association of Universities for Research in Astronomy, Incorporated, under NASA contract NAS5-26555. BMP acknowledges support from the US NSF through grant AST-1008882. GP acknowledges support via an EU Marie Curie Intra-European fellowship under contract no. FP-PEOPLE-2012-IEF- 331095. KCS acknowledges financial support from the Fondo Fortalecimiento de la Productividad Cient\'{i}fica VRIDT 2013. We acknowledge support by ISSI in Bern; the Netherlands Organization for Scientific Research;  the UK STFC; the French CNES, CNRS/PICS and CNRS/PNHE; the Swiss SNSF; the Italian INAF/PICS; the German Bundesministerium für Wirtschaft und Technologie/Deutsches Zentrum für Luft- und Raumfahrt (BMWI/DLR, FKZ 50 OR 1408).
\bibliographystyle{aa}
\bibliography{mybib.bib}
\end{document}